\documentclass[%
 reprint,
 amsmath,amssymb,
 aps,
prb,
floatfix,
superscriptaddress
]{revtex4-1}

\usepackage{graphicx}
\usepackage{bm}
\usepackage{braket}
\usepackage{hyperref}
\usepackage[usenames]{xcolor}
\usepackage[capitalize]{cleveref}
\usepackage{siunitx}
\usepackage{physics}

\def\black{\color{black}}

\begin{document}

\title{Chiral quantum optics using a topological resonator}

\author{Sabyasachi Barik}
\affiliation{Department of Physics, University of Maryland, College Park, MD 20742, USA}
\affiliation{Joint Quantum Institute, University of Maryland, College Park, MD 20742, USA and National Institute of Standards and Technology, Gaithersburg, MD 20899, USA}

\author{Aziz Karasahin}
\affiliation{Department of Electrical and Computer Engineering and Institute for Research in Electronics and Applied Physics, University of Maryland, College Park, Maryland 20742, USA}

\author{Sunil Mittal}
\affiliation{Joint Quantum Institute, University of Maryland, College Park, MD 20742, USA and National Institute of Standards and Technology, Gaithersburg, MD 20899, USA}
\affiliation{Department of Electrical and Computer Engineering and Institute for Research in Electronics and Applied Physics, University of Maryland, College Park, Maryland 20742, USA}

\author{Edo Waks}
\email{edowaks@umd.edu}
\affiliation{Department of Physics, University of Maryland, College Park, MD 20742, USA}
\affiliation{Joint Quantum Institute, University of Maryland, College Park, MD 20742, USA and National Institute of Standards and Technology, Gaithersburg, MD 20899, USA}
\affiliation{Department of Electrical and Computer Engineering and Institute for Research in Electronics and Applied Physics, University of Maryland, College Park, Maryland 20742, USA}

\author{Mohammad Hafezi}
\email{hafezi@umd.edu}
\affiliation{Department of Physics, University of Maryland, College Park, MD 20742, USA}
\affiliation{Joint Quantum Institute, University of Maryland, College Park, MD 20742, USA and National Institute of Standards and Technology, Gaithersburg, MD 20899, USA}
\affiliation{Department of Electrical and Computer Engineering and Institute for Research in Electronics and Applied Physics, University of Maryland, College Park, Maryland 20742, USA}

\begin{abstract}
Chiral nanophotonic components, such as waveguides and resonators coupled to quantum emitters, provide a fundamentally new approach to manipulate light-matter interactions. The recent emergence of topological photonics has provided a new paradigm to realize helical/chiral nanophotonic structures that are flexible in design and, at the same time, robust against sharp bends and disorder. Here we demonstrate  such a topologically protected chiral nanophotonic resonator that is strongly coupled to a solid-state quantum emitter. Specifically, we employ the valley-Hall effect in a photonic crystal to achieve topological edge states at an interface between two topologically distinct regions.  Our helical resonator supports two counter-propagating edge modes with opposite polarizations.  We first show chiral coupling between the topological resonator and the quantum emitter such that the emitter emits preferably into one of the counter-propagating edge modes depending upon its spin. Subsequently, we demonstrate strong coupling between the resonator and the quantum emitter using resonant Purcell enhancement in the emission intensity by a factor of 3.4.  Such chiral resonators could enable designing complex nanophotonic circuits for quantum information processing, and studying novel quantum many-body dynamics.
\end{abstract}
\maketitle
Chiral propagation of light can fundamentally alter the way it interacts with matter.   In particular, chiral light-matter interactions can control the directionality of spontaneous emission and modify photon-mediated interactions  between quantum emitters\cite{Lodahl2017}. These capabilities in-turn enable engineering of novel quantum states such as entangled spin states\cite{Pichler2015} and photonic clusters states \cite{Pichler2017}. Chiral light-matter interactions based on polarization-momentum locking of evanescent fields  have been achieved previously, for example, using optical fibers\cite{Petersen2014} and millimeter-scale bottle resonators\cite{Shomroni2014, Junge2013}. Chiral light-matter interactions have also been explored using purely opto-mechanical interactions \cite{Habraken2012, Peano2015, Vermersch2017, Calajo2019}. Recently, nanophotonics has emerged as a versatile platform to engineer chiral light-matter coupling in a compact and scalable fashion.  In particular, chiral/helical waveguides coupled to solid-state quantum emitters have demonstrated directional spontaneous emission \cite{Sollner2015, Coles2016}. However, extensions of these ideas to realize strong coupling between a chiral resonator and a solid-state emitter have remained elusive. This is mainly because of the challenges in designing a nanophotonic resonator while preserving polarization-momentum locking. 

Recently, topology has emerged as a new paradigm to design chiral photonic structures\cite{Ozawa2019}. In particular, an interface between two topologically distinct regions hosts edge states exhibit chiral/helical propagation of light where the photon’s momentum gets locked to a pseudospin degree of freedom, such as polarization. These edge states have the additional benefit that they are robust to deformations and disorders. Specifically, they also allow propagation of light around sharp bends and defects without scattering \cite{Wang2009, Hafezi2011, Rechtsman2013, Hafezi2013, Mittal2014, Cheng2016}, which is essential to engineer compact resonators that exhibit chirality\cite{Bahari2017,Bandres2018}.

Here, we demonstrate a topological resonator that exhibits strong light-matter interaction. We realize this resonator by creating an interface between two valley-Hall topological photonic crystals. By using an inhomogeneously broadened ensemble of quantum emitters as a broad-band light source, we first establish that our resonator supports edge modes that extend throughout the length of the resonator. Subsequently, using a single quantum emitter and a chiral waveguide coupled to the topological resonator, we demonstrate chiral spontaneous emission  where the direction of the emission depends on the polarization of the emitted light. Having established the chirality of our resonator, we finally proceed to demonstrate strong coupling between the quantum emitter and the topological resonator. Specifically, we use a magnetic field to tune a quantum emitter into resonance with the topological resonator and show Purcell enhancement of emission. Our results pave the way for explorations of strong interactions between multiple solid-state quantum emitters coupled via a chiral nanophotonic resonator. 

\begin{figure*}
\centering
\includegraphics[width=0.95\textwidth]{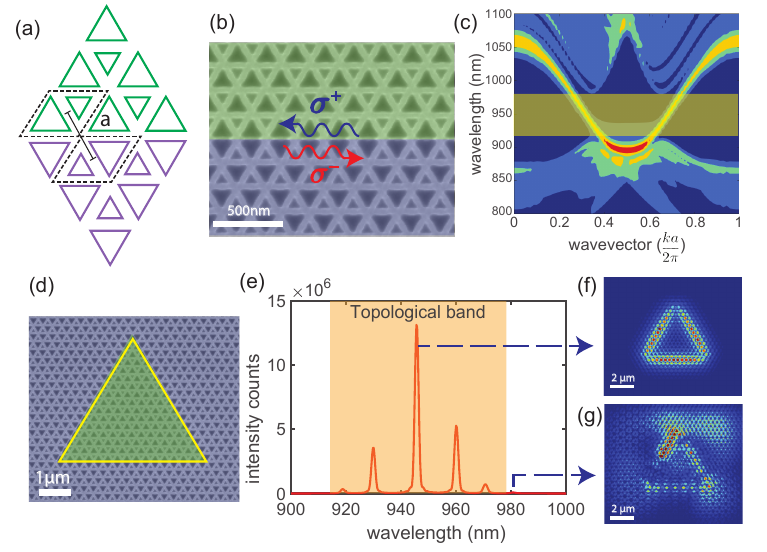}
\caption{Valley-Hall topological photonic crystal resonator and FDTD simulation results:  (a) Schematic of the topological interface showing two topologically distinct regions (violet and green), by interchanging the triangular holes sizes. Black dashed lines indicate the rhombic unit cells. (b) SEM image of the fabricated topological interface. (c) Simulated TE-band structure of the interface. The orange band highlights the edge band. (d) SEM image of a topological resonator shaped in the form of a super-triangle, with two topologically distinct regions (violet and green). (e) Simulated longitudinal modes of the resonator. As in (c), the shaded orange region corresponds to the topological edge band. (f) and (g) The electric field distribution of a resonator mode and a bulk mode.}
\end{figure*}

Our chiral resonator is based on a valley-Hall topological photonic crystal \cite{Dong2017,Chen2017, He2019, Noh2018, Shalaev2019, Ma2016}, composed of a honeycomb lattice of triangular holes with a lattice constant of ‘a’, as shown in Fig.1a. The two triangular holes in each rhombic unit cell (shown in red), have different sizes (\(\frac{1.3 a}{2 \sqrt{3}}\) and \(\frac{0.7 a}{2 \sqrt{3}}\)), which leads to the opening of topological bandgaps, at the K and K' points. Because of the time-reversal symmetry, the Berry curvature integrated over the entire Brillouin zone is zero. However, the Berry curvature at K and K' valleys have opposite signs. Interchanging the two triangular holes in the unit cell (green and violet regions) flips the sign of the Berry curvature at each valley. Therefore, by interfacing these two topologically distinct regions, we form a pair of counter-propagating edge states with opposite helicity at the boundary between the two topologically distinct regions (shown as an orange band in Fig.1c). These edge states are transverse electric (TE) modes composed of an in-plane electric field and can be selectively excited by placing an emitter at specific position along the interface of the waveguide with suitable circular polarization (see Appendix A)

The topological edge states are robust against sharp bends of 60 and 120 degrees, which preserve the symmetry of the structure. This robustness enables them to form a resonator using a super-triangle (Fig. 1d), where the inside/outside of the super-triangle corresponds to two topologically distinct regions. Analogous to the one-dimensional edge states, this confined resonator structure hosts two counter-propagating modes with opposite helicity. The 3D Finite-difference Time-domain (FDTD) simulation for a resonator of length $13 \mu m$ (50-unit cells) shows multiple longitudinal modes, separated by a free-spectral range (FSR) of $15 nm$ (Fig.1e). The calculated electric field profile of the resonator modes exhibits strong transverse confinement at the topological interface and extend over the super-triangle (Fig. 1f). Moreover, we do not observe any scattering at the three sharp corners of the resonator, indicating topological robustness. By contrast, the modes outside of the topological bandgap reside in the bulk and are not extended along the resonator perimeter (see Fig. 1g). Since these modes are not protected, their frequency and their field profile are susceptible to the disorder.   

\begin{figure}
\centering
\includegraphics[width=0.45\textwidth]{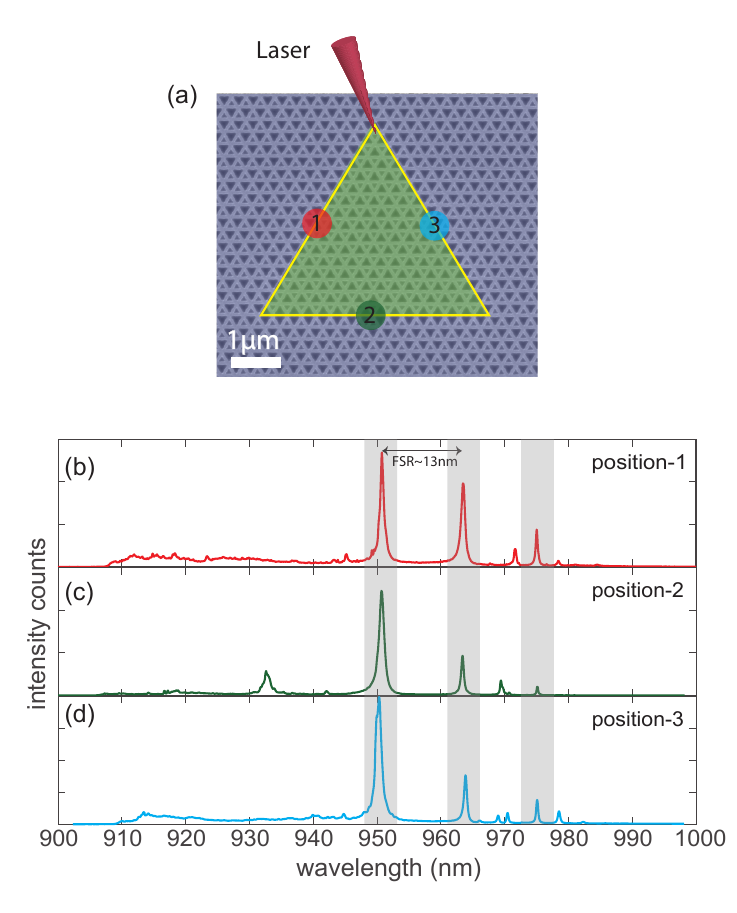}
\caption{ Topological waveguide-resonator system: (a) Schematic showing the excitation and collection points on a topological resonator device. Numbered marks show the position of collection spots. (b,c,d) The experimentally measured spectrum of a topological resonator from three different collection points along the length of the resonator. The peaks inside the grey regions correspond to the resonator modes. }
\end{figure}

We fabricated a resonator structure on a GaAs slab of thickness 160nm with an embedded layer of InAs quantum dots as quantum emitters. We patterned the structure using electron beam lithography followed by inductively coupled plasma reactive ion etching (see appendix B). We chose the lattice spacing of $a=$265nm, such that the topological band gap coincides with the emission spectra of quantum dots. All measurements were performed in a closed-cycle Helium cryostat, operating at 3K temperature. 

To show the modes of the topological resonator, we excite the quantum dots at the tip of the resonator (shown in Fig. 2a) using a high-numerical aperture objective and a continuous wave laser at 780nm. We use a high pump power of 100$\mu$W such that the quantum dots saturate and broaden to form a continuous internal white light source that probes the spectrum of the resonator. We collect the photoluminescence spectra from different points along the perimeter of the resonator (Fig. 2 b-d). We observe three peaks (within shaded grey regions) within the expected topological band which do not change when we move our collection point along the length of the resonator, indicating that these are the extended longitudinal modes of the topological resonator. Moreover, their free spectral range ($\approx$ 13nm) matches closely with the simulation. The other peaks in the PL spectra change with the collection point, and therefore, correspond to the bulk modes of the photonic crystal structure.

To probe the chiral coupling of an emitter to a topological resonator mode, we coupled the resonator to a helical topological waveguide, which is similarly designed (Fig. 3a). In this experiment, we excited the quantum dots at the point A on the resonator and collected the emission from the ends of the waveguide through the grating couplers. In order to analyze the chirality of the resonator mode, we numerically calculate the Poynting vector for the coupled resonator-waveguide device. Figure 3b shows the Poynting vector when the system is excited with a right circularly polarized dipole ($\sigma_-$). We observe that the electric field travels clockwise around the super-triangle, and then again chirally couples to the right of the waveguide. Due to time reversal symmetry, when the system is excited with a left circularly polarized dipole, we see the electric field travels anti-clockwise along the super-triangle, before exiting to the left of the waveguide (Fig. 3c). Note that the clockwise/anti-clockwise mode of the resonator couples to the right/left-going waveguide mode, respectively. This is due to the fact that the region below the waveguide has the same topology as the inside of the super-triangle, and therefore, the helicity is preserved at the boundaries. Here, we emphasize that the above mentioned polarization-momentum locking is dependent on the spatial position of the quantum dot with respect to topological interface.

\begin{figure}
\centering
\includegraphics[width=0.45\textwidth]{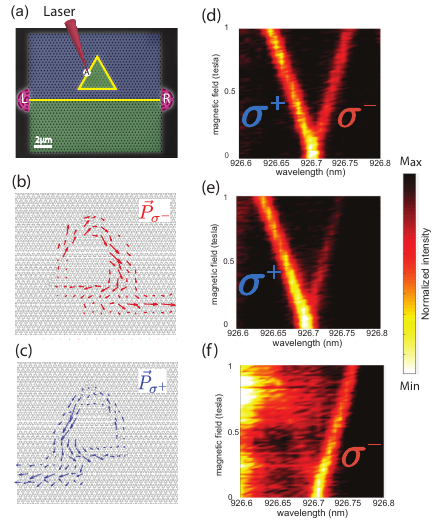}
\caption{Chiral waveguide-resonator-emitter coupling: (a) SEM image of the topological resonator coupled with a topological waveguide, terminated by two grating couplers (pink shaded half-circles). The yellow line indicates the interface between two topological regions, highlighted in green and violet. Point A indicates the point of excitation. (b,c) Simulated Poynting vector profile along the perimeter of the resonator-waveguide system, when the system is excited with a right circularly-polarized dipole and left circularly-polarized dipole respectively. (d,e,f) The measured PL signal, as a function of the magnetic field strength, collected from point A, Left grating, Right grating, respectively.}
\end{figure}

\begin{figure}
\centering
\includegraphics[width=0.45\textwidth]{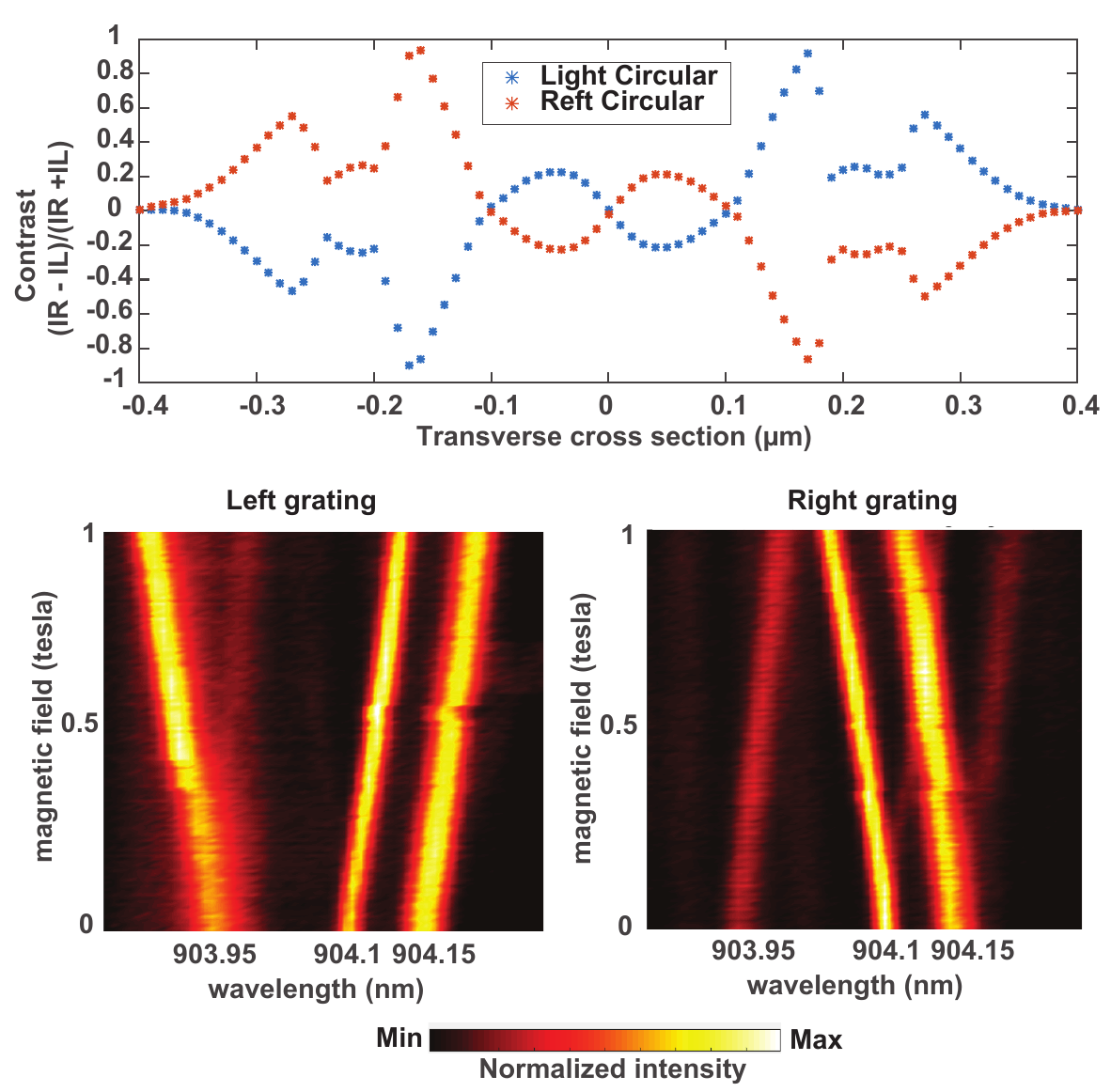}
\caption{(a)Simulated results for contrast between the two grating with two different types of polarization excitation along the transverse direction of the waveguide. (b),(c) The measured PL signal, as a function of the magnetic field strength, collected from the Left grating, and Right grating, respectively.}
\end{figure}

\begin{figure*}
\centering
\includegraphics[width=0.95\textwidth]{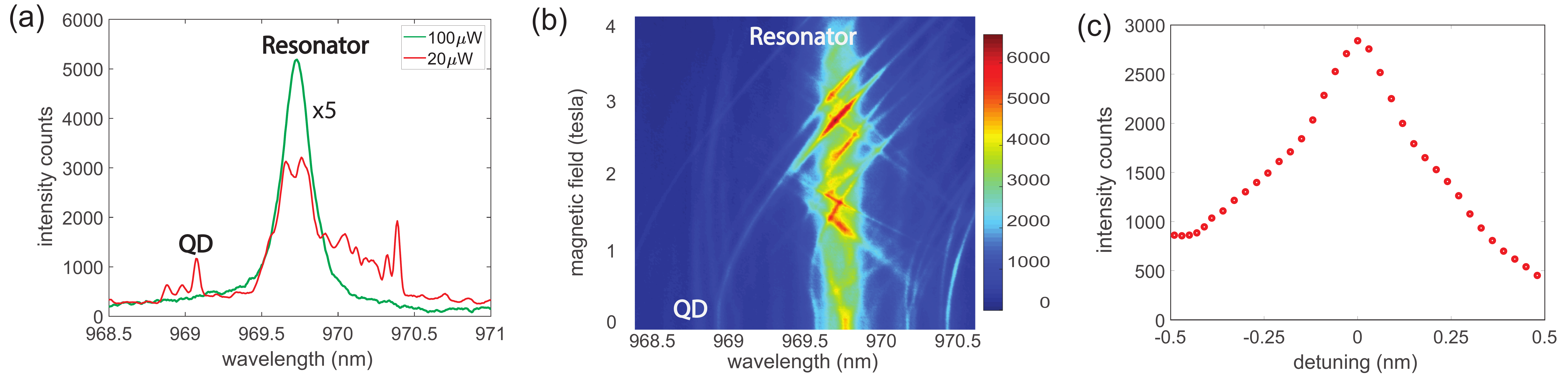}
\caption{Purcell enhancement of a quantum emitter coupled to a topological resonator: (a) The measured PL, at zero magnetic field, exhibiting the QD and the resonator mode at high and low excitation power. (b) Enhancement of emitter’s PL, when the QD is tuned to the topological resonator, by scanning the magnetic field. (c) The intensity of the QD’s emission, as a function of detuning from the resonator mode.}
\end{figure*}

To experimentally distinguish the chirality of the emitter-resonator-waveguide coupling, we perform a photoluminescence measurement on device shown in Fig. 3a. Specifically, we use two degenerate dipole transitions of the QD which are selectively coupled to the two helical edge states in the topological resonator.  We apply a magnetic field on QDs, in a Faraday configuration, such that the Zeeman shift lifts the degeneracy of the two transitions\cite{Bayer2002}.  This allows us to spectrally resolve the chiral nature of the light-matter interaction. Note that the helical nature of the resonator and the waveguide remains intact under such a magnetic field. We excite the sample at point A with a continuous wave 780nm wavelength laser at power 7$\mu$W, such that individual QDs can be spectrally resolved.  When the single is collected at the point A (Fig. 3d), we observe both branches of the Zeeman split QD spectrum, corresponding to two oppositely-circular polarization. However, when we collect the signal from either of the grating couplers (Fig. 3(e)-(f)), we observe a single branch, as a signature of chiral coupling. We calculate near 89$\%$  directionality between the two gratings in this measurement.

The above mentioned spin-momentum locking in this type of resonator system is dependent on the position of the quantum emitter along the waveguide. To affirm this fact, we excite a single resonator modes at different spatial points along the transverse direction with two circularly polarized dipoles and study the intensity variations at the two gratings. The results show that if the position of the emitting dipole is changed then the direction of the out-coupled light changes as shown in Fig. 4(a). Moreover, there are points of high chiral points where the magnitude of contrast reaches the maximum. Experimentally we further carry out PL measurements on another device. We find that within one excitation spot of the laser, depending on the position of the dot, the corresponding circular polarized emission will couple differently to the right/left propagating modes. This is apparent from Fig. 4(b) and (c) where the emission from a quantum dot at 903.95nm couples differently into the waveguide compared to the ones at 904.1nm and 904.15nm. Here, we want to note that even though the spin-momentum locking is similar to that found in ring resonators \cite{Lodahl2017,Shomroni2014,Junge2013}, these topological resonator modes have topological origin and arise from the fact that the two photonic crystals on either side of the triangular resonator have different band topology and therefore robust to bends and certain imperfection \cite{Shalaev2019}.  

\black In order to demonstrate that the quantum dots are coupled to the resonator, we measure their intensity as a function of cavity detuning. By scanning the magnetic field, one Zeeman branch of the QD spectrum can be tuned into the resonance of the topological resonator.  Figure 5a shows the spectrum at two different excitation power. At high excitation power (100$\mu$W), the saturated QDs emission reveals the cavity mode, as shown by the green curve.  By decreasing the excitation power to 20$\mu$W, the spectrum shows multiple individual quantum dot lines near the cavity resonance. Fig. 5b shows the measured spectra as a function of the magnetic field. As the magnetic field increases the quantum dots Zeeman split, and either the lower or upper branch crosses the cavity mode (depending on the initial detuning). The emission from the quantum dots is enhanced as they tune onto resonance with the cavity mode. 

We focus on one particular dot labeled ``QD''  in Fig. 5a.  This dot becomes resonant with the cavity at a magnetic field of 2.7 Tesla.  To quantify the degree of enhancement, we fit the quantum dot to a Lorentzian function at each magnetic field and plotted its intensity as a function of detuning from the cavity (Fig. 5c). At zero detuning from the resonator line, we see a nearly threefold increment in the count rates. From this emission enhancement, we estimate an intensity enhancement of 3.4.

In conclusion, we demonstrated a topologically robust and chiral interface between a photonic resonator and a quantum emitter.  This platform could provide a robust and scalable pathway to engineer chiral light-matter interaction between multiple emitters coupled to a single resonator. Such resonators could enable the generation of entangled states of photons, mediated by chiral coupling of photons to quantum emitters\cite{Lodahl2017}, such as superradiant\cite{Sipahigil2016,Bhatti2018} and cluster states\cite{Pichler2017}. On the other hand, one could conceive generating entangled states of several solid-state spins, mediated by the helical-circulating photonic modes\cite{Pichler2015}. In contrast to conventional waveguides, the mediated interaction strength between spins does not depend on the distance, since the emitters cannot form a mirror in a chiral interface. Ultimately, chiral and topological interfaces provide a new approach to study QED in a new regime \cite{Hafezi2013njp, Umucalilar2012, Bello2018, Perczel2017}.

\section*{ACKNOWLEDGMENTS}
This work is supported by NSF 1820938, 1430094 Physics Frontier Center at the Joint Quantum Institute and the Air Force Office of Scientific Research–Multidisciplinary University Research Initiative (FA9550-16-1-0323). 

\section*{APPENDIX A : The polarization profile of a topological resonator }

The electric field for the topological resonator mode is composed of two circularly-polarized components, where the high-field intensity points appear in different locations, as shown in Fig.6 Although these waveguides exhibit polarization-momentum locking, these plots indicate that the polarization profile changes in the transverse direction. If an emitter is located at the peak of right-circularly polarized light, and the QD is prepared to emit into right/left-circularly polarized light, then the emitted light travels in a clockwise/anti-clockwise fashion around the resonator.

\begin{figure}[h]
\includegraphics[width=0.9\columnwidth]{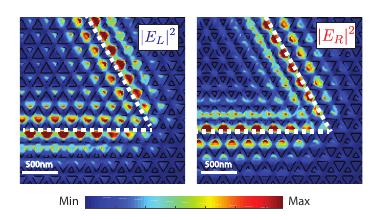}
\caption{Simulated left/right circular polarization components of the electric field. White dotted line marks the perimeter of the resonator.}
\end{figure}

\section*{APPENDIX B : Sample fabrication}
The initial wafer is composed of 160 nm GaAs membrane with quantum dots grown at the center of the growth axis. 1 $\mu$m sacrificial layer of Al\textsubscript{0.8}Ga\textsubscript{0.2}As beneath the active layer is used to undercut and to create suspended structures. The quantum dot density was 50 $\mu m ^{-2}$, which is high enough to find emitters resonant with the cavity mode. Based on the given quantum dot density and inhomogeneous broadening, the photon emitted by one quantum dot to be scattered by a second is calculated to be extremely unlikely.
Samples are first spin-coated with ZEP-520A positive e-beam resist, followed by patterning using 100 keV high-resolution e-beam system. Exposed regions are developed by using ZED50 developer. After patterning, chlorine-based directional ICP etching is performed to transfer the patterns from ZEP520A hard mask to the GaAs layer. Lastly, selective wet etching using Hydrofluoric acid is performed to create a suspended structure with air on top and bottom.

Due to imperfections in e-beam exposure and directional dry etching, fabrication of small and sharp corners is challenging. The loss in the resolution is mitigated by using a modified layout design for triangles. We incorporate rectangular structures in the periphery in order to facilitate a homogeneous undercut of the sacrificial layer. Also, we fabricated an array of devices with different e-beam doses. 

\section*{APPENDIX C :Setup}
\begin{figure}
\includegraphics[width=0.95\columnwidth]{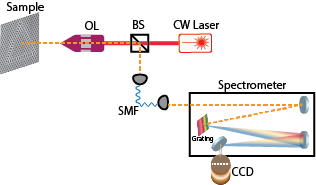}
\caption{Schematic of the experimental setup. OL, BS, SMF represent objective lens, beam splitter, single mode fiber,  respectively.}
\end{figure}

For photoluminescence measurement, we mount the sample on the cold finger of a cryostat (Attocube-Attodry system) which can be cooled down to a temperature 3K. The cold mount is surrounded by a superconducting magnet, which provided a variable magnetic field ranging from zero to 6 Tesla. In our setup, the sample is mounted in a Faraday configuration, i.e., the applied magnetic field is perpendicular to the sample plane. 

We use a confocal microscope setup (Fig. 7) for both exciting QDs and also collecting the photoluminescence (PL) signal from them. We use a continuous-wave 780nm diode laser for excitation. In the experiment, the collimated laser beam is focused on the sample by an objective lens with NA=0.8. We also image the whole sample surface by shining a broadband light. This helps us to locate both the excitation and collection spots on the sample. The collected light is then focused on a single mode fiber (SMF) which acts as a spatial filter. For spectrally resolving individual QD lines, we pass this collected light through a spectrograph fitted with grating. In our case the spectral resolution is 0.02nm. This resolved light is then focused onto a  nitrogen-cooled charge-coupled device (CCD) camera array capable of imaging at the single-photon level.

In experiment, due to 0.6 $\mu m ^2$ spot size of the laser spot, we excite roughly 25 quantum dots at the same time. Now owing to their inhomogeneous broadening and random placement all over the structure, they emit at different wavelengths and couple differently into the waveguide mode. We then isolate individual dots by spectrally filtering them through a grating spectrometer. We then confirm their coupling into the waveguide by spectrally locating them from emissions from different places of the device e.g. either from different locations on the resonator, or gratings at the end of the waveguide.

As mentioned in the main text, we used different measurement schemes for obtaining results depicted in different figures in the main text. We performed these measurements on different types of devices with different does array. Data presented in Fig.2 , 3, 4 has been taken from sample $\#$ D240\_m3s2b\_22 ,  $\#$ D240\_m3s2b1d3\_22,  $\#$  D240\_m3s2b1\_21 respectively.

\section*{APPENDIX D : The quality factor of resonator in hybrid structures}

\begin{figure}
\includegraphics[width=0.95\columnwidth]{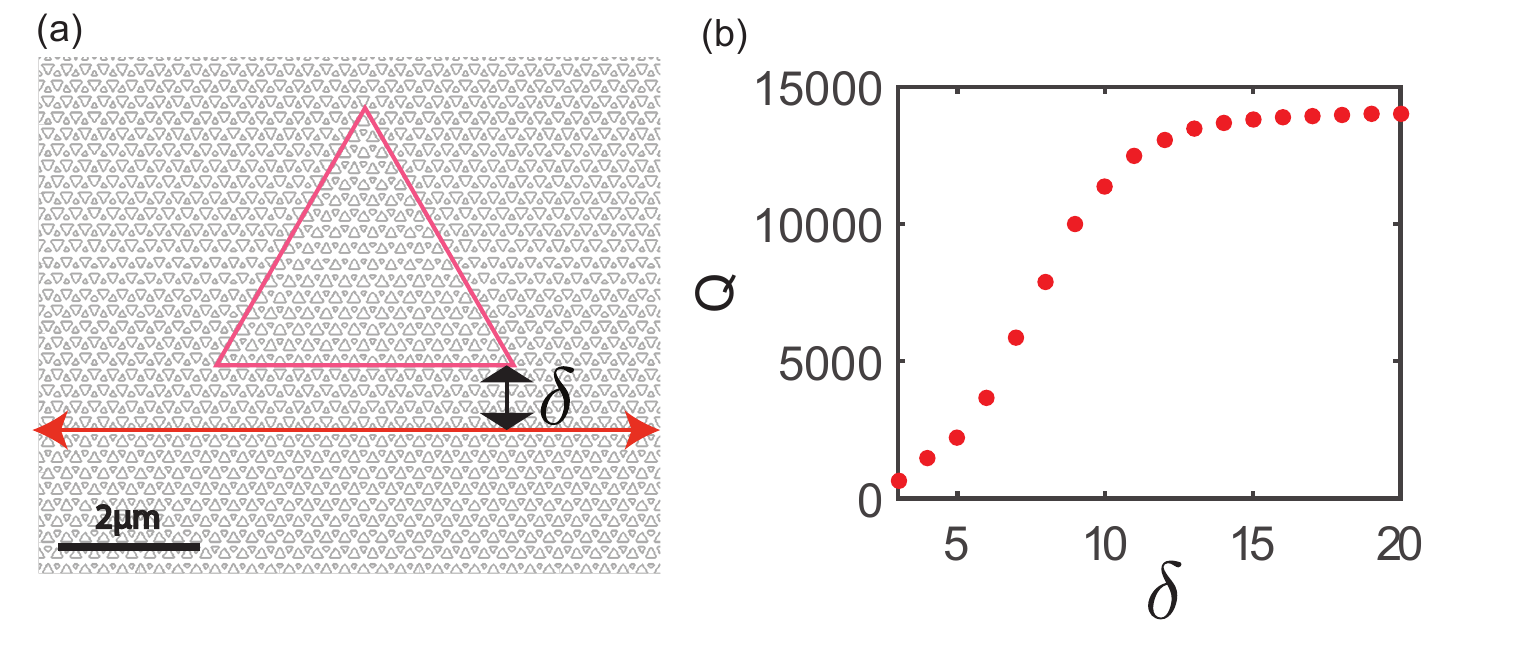}
\caption{(a)Schematic of the simulation setup. Red super triangle indicates the resonator perimeter. The double arrowed red line indicates the position of the waveguide. (b) The quality factor (Q) versus distance ($\delta$), showing saturation of Q at high values of  $\delta$ }
\end{figure}

We simulated the coupled waveguide-resonator system for different system parameters. In particular, we look at the behavior of resonator quality factor (Q) as we vary the amount of coupling between the waveguide and the resonator. Since we fix the perimeter of the resonator, the coupling strength only depends on the distance between the side of the resonator (facing the waveguide) and the topological waveguide. We define the parameter $\delta$ as the number of unit cells, characterizing this coupling distance (Fig.8(a)). As shown in Fig. 8(b) the Q of the resonator attains a saturation after $\delta$. In other words, at this parameter, the resonator is weakly coupled into the waveguide. Also depending on the different values of $\delta$ one can create resonators of different Q. In our experiment, we choose $\delta$ to be 6.  

\begin{figure*}
\includegraphics[width=0.8\textwidth]{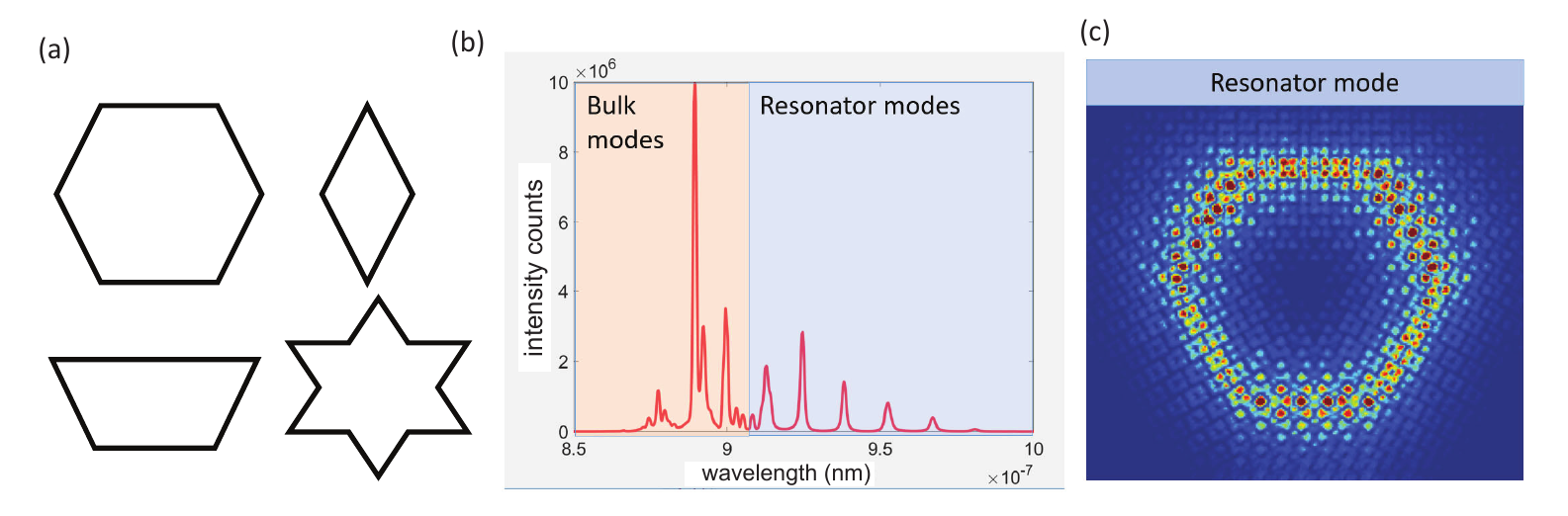}
\caption{(a) Possible resonator shapes using valley-Hall edge states (b) Simulated longitudinal modes of a hexagonal resonator showing both bulk modes and resonating modes. (c) The in-plane electric field distribution for one of the resonator modes.  }
\end{figure*}

\section*{APPENDIX E : Resonator designs with Valley-Hall physics}

Apart from a triangular geometry, one can also fabricate other types of resonator designs with the valley-hall topological photonic crystal. Since these valley edge modes are immune to 60\textsuperscript{0} and 120\textsuperscript{0}, one can think of other resonator designs such as hexagon, rhombus, 3-side equal trapezoid, or more complicated designs, such as a hexagram. 

Here we analyze a hexagonal resonator. As in the case of the triangular resonator in the main text, here the inside region of the hexagon is topologically distinct from the outside zone. Fig. 9(b) shows the simulated modes in such a system. Color shaded regions correspond to different regimes on the band structure. Specifically, the orange shade represents the bulk modes in the system which are susceptible to disorder thus can provide random localized modes. And the blue band corresponds to edge band of the band structure. Thus these modes show strong confinement along the perimeter of the resonator (Fig. 9(c)).

\bibliographystyle{apsrev4-1}
\bibliography{topo_resonator}
\end{document}